# Mobility Impact on Performance of Mobile Grids

A. S. Nandeppanavar[1], M. N. Birje[2], S. S. Manvi[3], Shridhar[4]
[1, 2]Department of ISE, Basaveshwar Engineering College, Bagalkot-587102, India
[3]Department of E&CE, Reva Institute of Technology & Management, Bangalore, India
[4]Department of E&CE, Basaveshwar Engineering College, Bagalkot-587102, India

*Abstract*— Wireless mobile grids are one of the emerging grid types, which help to pool the resources of several willing and cooperative mobile devices to resolve a computationally intensive task. The mobile grids exhibit stronger challenges like mobility management of devices, providing transparent access to grid resources, task management and handling of limited resources so that resources are shared efficiently. Task execution on these devices should not be affected by their mobility.

The proposed work presents performance evaluation of wireless mobile grid using normal walk mobility model. The normal walk model represents daily motion of users and the direction of motion is mostly symmetric in a real life environment; thus it is effective in location updating of a mobile station and in turn helps task distribution among these available mobile stations. Some of the performance parameters such as Task Execution Time, task failure rate, communication overhead on Brokering Server and Monitoring Cost are discussed.

## I. INTRODUCTION

Mobile grid computing is one of the emerging grid types, with two underlying technologies: Mobile Computing and Grid Computing. Mobile grids focus on resource sharing among wireless mobile devices for distributed applications and are characterized by relatively high mobility and limited resources. Common resources shared are: computing power, data storage/network file system, communication and bandwidth, application software. These mobile grids exhibit stronger challenges like mobility management of devices, providing transparent access to grid resources, and handling of limited resources so that resources are shared efficiently. Therefore it is necessary to develop an environment model which can represent the issues mentioned above.

The work given in [1] focuses on wireless mobile grid for collaborative problem solving considering mobility related issues such as effect of mobility on performance of grid and network instability due to mobility. This mobile grid framework allows mobile devices to work collaboratively on computationally expensive tasks. Such a task is decomposed into smaller tasks and distributed across the other mobile devices willing to share their computational power with others.

Mobility models are important parameters for location updating of Mobile Stations (MS). We consider Normal Walk Mobility Model [2] to represent mobility pattern of users and to decide their location. The work also focuses on finding the direction of movement of MS and to predict instantaneously the Base Station Controller (BSC) with which handover occurs. Performance is analyzed based on parameters like task execution time, task failure rate, communication overhead on Brokering Server (BS) and monitoring cost.

The paper is organized as follows: Section II presents the related works on wireless mobile grid architectures, mobility models, mobility management and handovers in wireless mobile networks. Proposed work is described in section III. Section IV discusses about the results. And section V concludes the work.

## II. RELATED WORKS

Ian Foster et al have defined grid [3] as "flexible, secure, coordinated resource sharing among dynamic collection of individuals, institutions, and resources what we refer to as virtual organizations". They have highlighted the need for grid technology in virtual organization.

T. Phan et al [4] have presented the challenge of integrating mobile devices with computational grid. The integration is provided through the use of an Interlocutor, which acts as proxy for cluster of Minions. There is no selection strategy to replace an interlocutor which has moved to another cell.

Kurkovsky et al [1], [5], and [6] have proposed an agent based approach to the design of wireless grid architecture to solve computationally expensive tasks. This architecture enables mobile devices within a wireless cell to form computational grid. It has several limitations, one of which being the inadequate consideration for the mobility of the mobile agents. Tasks are indiscriminately aborted by Subordinates and/or Initiators whenever these mobile agents move to neighboring cells.

P. Mudali et al [7] have proposed an extension to the architecture by Kurkovsky et al in [1]. They have proposed a multi-cell wireless computational grid, which is based on location area concept in GSM cellular networks. The proposed wireless computational gird is capable of greater device mobility tolerance than proposed in [1]. But still there is need to introduce mobility management schemes in proposed architecture.

Ian F. Akyildiz [8] et al have proposed a simplified random walk model for hexagonal cell configuration where, probability states gives performance of the model. Further Guoliang Xue [9] and Md. Imdadul Islam and A.B.M. Siddique Hossain [10] have proposed improved models by reducing number of probability states to improve the performance.

Chiu-Ching Tuan et al have proposed a novel normal walk model for PCS networks with mesh cell configuration in [11] and compact normal walk model for hexagonal cell configuration in [2]. They represent the daily mobility behavior of an individual mobile station that moves from a cell to another, in PCS networks.

Jingyuan Zhang has described various location management schemes and mobility modeling method used for cellular networks in [12], and also provided the comparison of these methods.

M. N. Birje [13] et al have proposed a prediction based handover model for multiclass traffic in wireless mobile networks by using software agents, considering two cases: local handoff (between BSC's connected to same mobile switching center (MSC)), and global handoff (between BSC's connected to different MSC).

### III. PROPOSED WORK

In this section we describe the architecture used in proposed work, mobility modeling, and location updating mechanisms.

*A. Architecture*

Figure 1 shows considered architecture. It includes two Virtual Organizations spanning over different Actual Organizations, which are monitored and controlled by a BSC. Its components are described briefly as follows:

Virtual Organization (VO): is the one which spans multiple actual organizations and transcends greater amount of geographical, organizational, and other type's related to intellectual property rights and national laws.

Actual Organization (AO): represents a single organization in a single place.

Base Station Controller (BSC): provides service to number of AOs through BTS. It has two servers BS and BSMS.

Base Transceiver Station (BTS): it supports the communication between BSC and mobile stations within AO. The BTS is fixed and is able to communicate with mobile stations using its radio transceiver.

Mobile Station (MS): is nothing but the mobile node which the user is holding. MS within an AO can play two roles: Initiator and Subordinator. Any device which is ready to take part in problem solving in grid environment is called as Subordinator. Initiator is the one which initiates or requests a task to be solved. Any subordinate may become an Initiator of distributed task if its user requests a large and/or computationally intensive task to be solved. In this case initiator is responsible for submitting such a distributed task to Brokering Server.

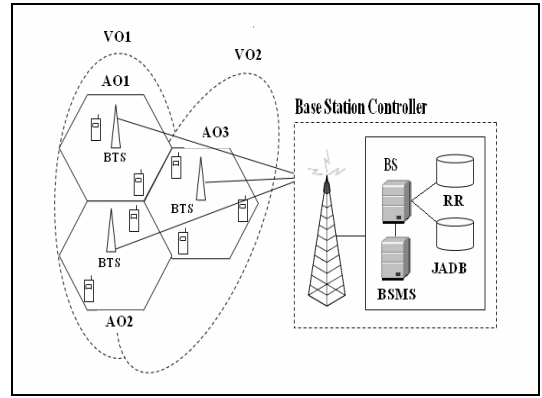

Figure 1. Architecture

Brokering Server (BS): it has the responsibility of task distribution and coordination in solving computationally intensive tasks. It knows the initial residing status of all devices in VO. Task distribution is based on available resources. BS has two data stores as below:

Resource Repository (RR): keeps track of available resources. Details of currently available resources of all mobile stations are stored in Resource Repository.

Job Allocation Database (JADB): keeps track of job distribution during task execution. Information about all subtasks allotted to different mobile stations and also results after performing operation are stored in this data base.

Base Station Monitoring Server (BSMS): It supports communication among mobile stations in wireless grid. It also keeps track of all the mobile stations available in the wireless grid. Each node entering or leaving the wireless grid should inform Base Station Monitoring Server. It maintains information about the mobile station's current location, which in turn helps to predict handoff in advance.

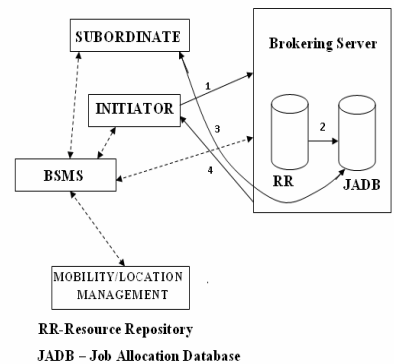

Figure 2. Interaction Diagram

Figure 2 explains the interaction among different components of the architecture. (1) Initiator sends the job to be solved to the BS; BS looks for eligible nodes in RR, so that sub jobs can be distributed to them. (2) Details of sub job and nodes to which jobs are to be distributed are stored in JADB. (3) BS distributes sub jobs to eligible subordinates in the grid. Subordinates solve the sub jobs and return their partial results to the BS, which is stored in JADB (4) the initiator collects the partial results from the BS. BSMS supports the communication among all the nodes and servers. It is also responsible for location and mobility management of the mobile node.

### B. Mobility modeling and location updating

Proposed architecture considers Normal Walk mobility model described in [11] for modeling movements of mobile stations within Actual Organizations. Normal Walk mobility model is a multi-scale, straight-oriented, mobility model. It represents the daily mobility patterns of a mobile station and the direction of motion is mostly symmetric in a real life environment. The mobile station moves in one step. Each move is based on previous move and is obtained by rotating previous move by an angle of θ in anticlockwise. This angle is called as moving angle, and helps to determine the next relative direction in which an MS (mobile station) moves across a cell in single step.

The drift angle θ in this model is a continuous random variable and its value helps to decide angle of movement of mobile station. The probability distribution of θ is assumed to approach normality rather than randomization with two parameters:

*Mean (μ)* with zero-degree.

*Standard Deviation (σ)* in the interval [5°, 90°]

Varying σ can redistribute the probabilities associated with θ and in turn helps to make the movement patterns more realistic to represent the user mobility. Thus any movement yielded from this model is function of σ, and is called a normal walk. The normal distribution of θ is represented as:

$$\theta \sim N(0°, \sigma^2) \quad (1)$$

And probability density function of θ is defined by:

$$f(\theta) = \frac{1}{\sqrt{2\pi}\sigma} \cdot e^{-\frac{1}{2}\left(\frac{\theta}{\sigma}\right)^2}, -270 < \theta < 270 \quad (2)$$

This drift angle θ also helps to determine one of the six relative directions in which an MS handoffs from a hexagonal cell (in figure 3) to another in next step. The six directions are indexed by 'k' and given below:

- Making U-turn or turning back (B, k=0).
- Turning right (R, k=1).
- Moving front-right (Fr, k=2)
- Moving front or forward (F, k=3).
- Moving front-left (Fl, k=4)
- Turning left (L, k=5).

The direction k is not absolute for a mesh cell, but is relative to the inlet that an MS is currently visiting. The range of each direction k is confined to lie between two fixed angles. The confining angles are calculated as:

$$angF = \tan^{-1}\left(\frac{Ro}{4Ri}\right)$$

$$angFl = \tan^{-1}\left(\frac{Ro}{Ri}\right)$$

$$angL = \tan^{-1}(\infty) \quad (3)$$

where Ri and Ro represent inner and outer radii of hexagonal cell respectively. Thus $angF = 16.1°$, $angFl = 49.1°$ and $angL = 90°$.

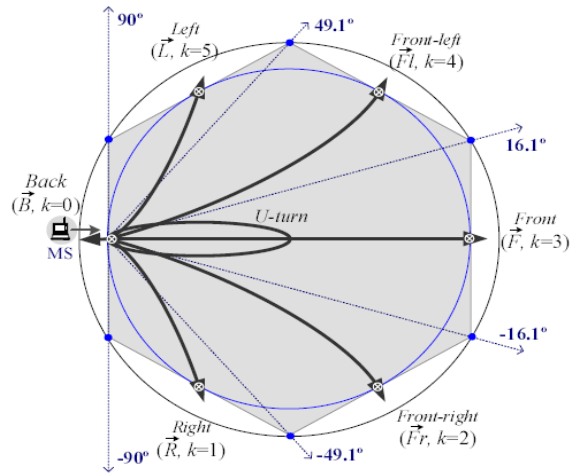

Figure 3. Hexagonal Cell

To find the probability of each of the six directions at next step, θ is standardized into Z (a continuous random variable), to find cumulative probabilities. Let Z= θ/σ such that Z has a standard normal distribution, Z ~ N (0, 1). The probability distribution function (pdf) and cumulative distribution function (cdf) are respectively given by

$$\varphi(z) = \frac{1}{\sqrt{2\pi}} \cdot e^{-z^2/2}$$

$$\phi(z) = \int_{-\infty}^{z} \varphi(\omega) d\omega \quad (4)$$

Considering the equations (3), (4) the six probabilities associated with each direction k, with given σ are given by:

$$f = 2 \cdot \phi\left(\frac{angF}{\sigma}\right)$$

$$l = \phi(\frac{angL}{\sigma}) - \phi(\frac{angFl}{\sigma})$$

$$fl = \phi(\frac{angFl}{\sigma}) - \phi(\frac{angF}{\sigma})$$

$$b = 1 - 2\phi(\frac{angL}{\sigma})$$

$$fr = fl$$

$$r = l \quad (5)$$

Movement of mobile node follows normal walk model, where BS knows the initial location of the device. Location updating of the device during movement follows the algorithm below:

1. Find the new location of the MS based on its current location and moving angle

2. If the MS moves out of VO, but within same AO, then information update is done to reflect its new location

3. If the MS moves out of VO, and also from local AO, then handover occurs

*C. Job Distribution and Execution*

BS of the proposed architecture is responsible for job handling, so that sub jobs of a given large job are distributed and partial results are collected back. BS has a data base called as JADB to keep track of job distribution among mobile devices within a wireless grid. It includes sub jobs distributed, identity of node to which jobs allotted and partial results. If a mobile device within wireless grid initiates a new job, the source code needed to run the corresponding job and all relevant parameters are submitted to the BS. BS generates sub jobs and assigns the sub jobs to available eligible subordinates, and waits to collect back the partial results.

Steps used to distribute and solve a computationally expensive job in the wireless grid environment can be described as below:

1. The user of a mobile device initiates a computationally intensive job

2. The Initiator creates sub jobs and transmits them to Brokering Service

3. The Brokering Service stores the received sub jobs in the JADB

4. The Brokering Service uses heuristics to find a subordinate in the RR for each sub job

5. The Brokering Service transmits sub jobs to the Subordinates

6. Each chosen Subordinate receives a sub job

7. Subordinates execute the sub jobs. During job execution

    a) if the subordinate moves out of VO, but within same AO, then the task assigned to that subordinate is continued to execute on it and the location of the subordinate is updated

    b) if the subordinate moves out of VO, and also from local AO, then task assigned to it is terminated. BS then redistributes it to another eligible subordinate and updates its JADB.

8. The Brokering Service receives partial results from subordinates that have finished their sub jobs and store them in the JADB

9. The initiator may chose to accept partial results or wait until all results are received from Brokering Service

The efficiency of the proposed work is measured using the average time taken for a task to be initiated, distributed, solved by the subordinates and returned back to the initiator. This average task execution time depends on different parameters like population of wireless grid and device mobility. Population of wireless grid specifies number of mobile nodes present in the wireless grid. Device mobility refers to the probability of a new device joining a wireless grid or a current device leaving the wireless grid per unit of time.

Some of the performance parameters evaluated are as follows:

**Task execution time:** It is the time taken for a task to be distributed by initiator, solved by the subordinates, and returned back to the initiator. It depends on different parameters like mobility factor (percentage of mobile nodes joining or leaving the wireless grid) of nodes within grid environment and grid population. To study the effect of grid population on execution time, the simulator is run considering different grid population. And to study the effect of mobility on execution time, the simulator is run under two scenarios: with and without mobility of nodes. In without mobility, nodes are considered to be at fixed location and tasks are distributed among them to find execution time. Where as in mobility, number of nodes is assumed fixed and mobility factor is varied.

**Location Monitoring overhead:** Location monitoring is keeping track of nodes status. BSMS has the responsibility of keeping track of nodes. It updates the database whenever a node enters or leaves the grid. Thus number of updates helps to determine the location monitoring overhead. The wireless grid simulator is run with different mobility factors and number of updates for each run is tabulated.

**Communication overhead:** location updating overhead leads to communication overhead because, location updating increases number of communications among BS and BSMS. This in turn leads to increase in bandwidth utilization. Bandwidth utilization for different mobility factors is recorded to study its effect on communication.

## IV. SIMULATION AND RESULTS

The proposed model has been simulated for various wireless grid scenarios. It considers *m* number of VOs controlled by a single BSC, with *c* number of AOs in each VO. The number of nodes in a VO varies between *n1* to *n2*. Each VO is allocated *v* Mbps bandwidth. Mobility factor in any VO is represented by *mf*.

The inputs considered for simulation are: *m* = 2, *c* = 2 for both VOs, *n1* = 30 and *n2* = 90. *v* = 40. *mf* = varies between 10% to 40%. Results are depicted using graphs as below:

Figure 4 shows the execution time calculated considering different grid size (grid population), where mobile nodes are stationary. The graph depicts that increase in grid size reduces the execution time i.e. increasing grid size in turn helps to provide more resources for executing the tasks and thus can be executed faster with more resources.

Figure 5 shows effect of mobility on execution time considering different mobility factors. The graph depicts that execution time gradually increases with an increase in mobility factor, because higher mobility factor leads to higher rate of task abortion and reallocation of task.

Figure 6 shows the effect of mobility factor on number of updates. The graph depicts that location update rate is more as mobility is more. Increasing mobility factor lead to increase in number of updates i.e. location update cost.

Figure 7 shows the effect of mobility factor on bandwidth utilization. The graph depicts that bandwidth utilization rate is more as mobility is more.

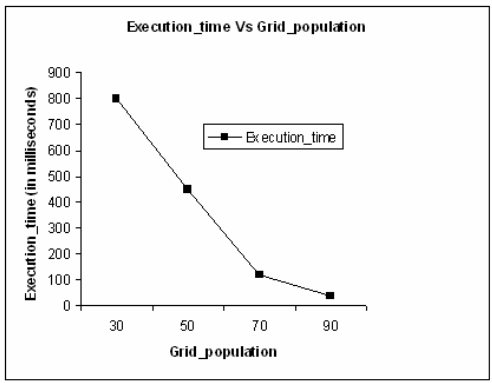

Figure 4. Execution time Vs. Grid_population

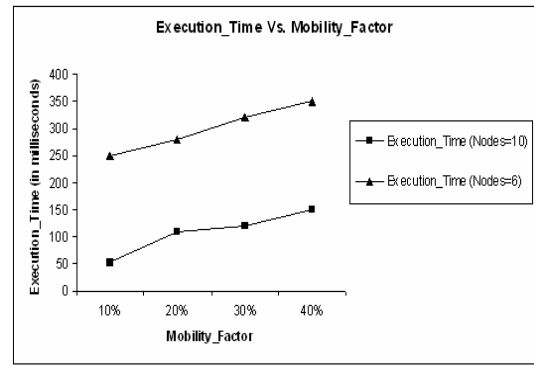

Figure 5. Execution time Vs. Mobility_factor

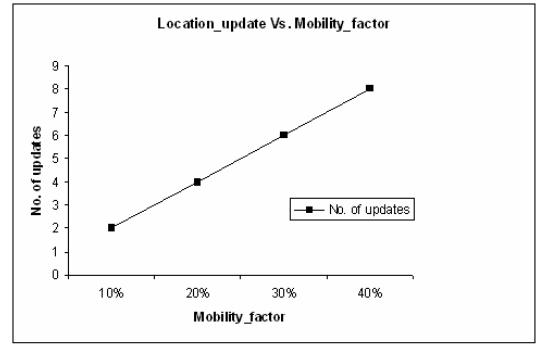

Figure 6. No. of updates Vs. Mobility_factor

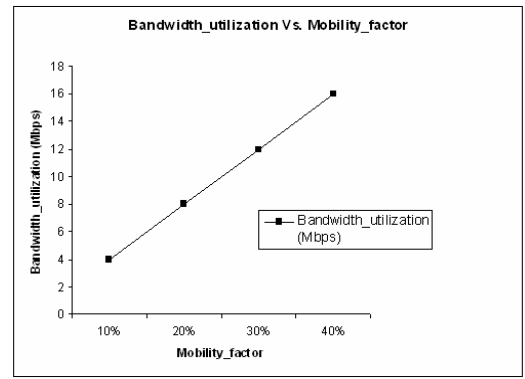

Figure 7. Bandwidth_utilization Vs. Mobility_factor

## V. CONCLUSION

The proposed work presents implementation of normal walk mobility model in a wireless mobile grid environment for a Virtual Organization with two actual organizations. Normal walk mobility model helps to decide the next relative direction in which mobile station moves and in which it handoffs form one AO to another. This model in turn helps BSMS to keep track of number of mobile stations available in AOs; so that Brokering Server can distribute tasks among available nodes for tasks request sent to Brokering Server.

The proposed work considers Normal walk model because, it represents daily motion of users. The directions of motion are mostly symmetric in a real life environment, thus it is effective in location updating of a MS. The work also shows the effect of grid population and mobility factor on execution time and monitoring cost. It can be concluded that, high rate of resource availability decreases task execution time and higher mobility factor increases execution time. Finally increase in mobility factor increases monitoring overhead.

AUTHORS PROFILE

**Anupama S. Nandeppanavar** received B.E. in ISE from BLDEA's college of Engineering, Bijapur in 2005, and now she is PG student in Basaveshwar Engineering College, Bagalkot. Currently she is working as Lecturer Basaveshwar Engineering College, Bagalkot. Her interested area is computer networking.

**Mahantesh N. Birje** received B.E. in CSE from UBDT college of Engineering, Davangere in 1997, M.Tech. in CSE from Basaveshwar Engineering College, Bagalkot in 2005, and currently he is the research scholar at Visvesvaraya Techological University, Belgaum. He is working as Asst. Professor in the department of Information Science and Engineering, Basaveshwar Engineering College, Bagalkot Karnataka, INDIA. His area of interest include Grid computing, Multimedia Communications, and Agent technology. He has published 4 refereed journal papers, and 7 international conference papers.

**Sunilkumar S. Manvi** received M.E. degree in Electronics from the University of Visweshwariah College of Engineering, Bangalore, Ph.D degree in Electrical Communication Engineering, Indian Institute of Science, Bangalore, India. He is currently working as a Professor and Head of Department of Electronics & Communication Engineering, REVA Institute of Technology and Management, Bangalore, India. He is involved in research of Agent based applications in Multimedia Communications, Grid computing, Ad-hoc networks, E-commerce and Mobile computing. He has published 3 books, 3 book chapters, 25 refereed journal papers, and about 75 refereed conference papers. He has given many invited lectures and has conducted several workshops/seminars/conferences.

**Shridhar** received B.E. in E & CE from Gulbarga University in 1988, M.Tech. in digital electronics & advanced communications from KREC, Surathkal in 2000. He is currently working as Asst. Professor in the department of Electronics and Communications Engineering, Basaveshwar Engineering College, Bagalkot. His area of interest include signal processing and control systems.